\documentclass[aps,floats]{revtex4}
\usepackage{amsmath}
\usepackage{graphicx,epsfig}

\begin{document}
\bibliographystyle {plain}

\def\oppropto{\mathop{\propto}} 
\def\opsimeq{\mathop{\simeq}}
\def\opoverderline{\mathop{\overline}}
\def\operarrow{\mathop{\longrightarrow}}
\def\opsim{\mathop{\sim}}

\def\fig#1#2{\includegraphics[height=#1]{#2}}
\def\figx#1#2{\includegraphics[width=#1]{#2}}


\title{ Freezing transition of the directed polymer in a $1+d$ random
medium  : \\ 
location of the critical temperature and unusual critical properties } 

\author{ C\'ecile Monthus and Thomas Garel}
 \affiliation{Service de Physique Th\'{e}orique, CEA/DSM/SPhT\\
Unit\'e de recherche associ\'ee au CNRS\\
91191 Gif-sur-Yvette cedex, France}

\begin{abstract}

\bigskip

In dimension $d \geq 3$, the directed polymer in a random medium
undergoes a phase transition between a free phase at high temperature
and a low temperature disorder dominated phase. For the latter phase,
Fisher and Huse have proposed a droplet theory based on  the scaling
of the free energy fluctuations $\Delta F(l) \sim l^{\theta}$ at scale
$l$. On the other hand, in related growth models belonging to the KPZ
universality class, Forrest and Tang have found that the height-height
correlation function is logarithmic at the transition. For the
directed polymer model at criticality, this translates  into
logarithmic free energy fluctuations $\Delta F_{T_c}(l) \sim (\ln
l)^{\sigma}$ with $\sigma=1/2$. In this paper, we propose a droplet
scaling analysis exactly at criticality based on this logarithmic
scaling. Our main conclusion is that the typical correlation length
$\xi(T)$ of the low temperature phase, diverges as $ \ln \xi(T) \sim (
- \ln (T_c-T) )^{1/\sigma} \sim ( - \ln (T_c-T) )^{2}  $, instead of
the usual power-law $\xi(T) \sim (T_c-T)^{-\nu}$. Furthermore, the
logarithmic dependence of $\Delta F_{T_c}(l)$ leads to the conclusion
that the critical temperature $T_c$ actually coincides with the
explicit upper bound $T_2$ derived by Derrida and coworkers, where
$T_2$ corresponds to the temperature below which the ratio
$\overline{Z_L^2}/(\overline{Z_L})^2$ diverges exponentially in
$L$. Finally, since the Fisher-Huse droplet theory was initially
introduced for the spin-glass phase, we briefly mention the
similarities and differences with the directed polymer model. If one
speculates that the free energy of droplet excitations for
spin-glasses is also logarithmic at $T_c$, one obtains a logarithmic
decay for the mean square correlation function at criticality
$\overline{ C^2(r) } \sim 1/\left( \ln r  \right)^{\sigma}$, instead
of the usual power-law $1/r^{d-2+\eta}$.

\end{abstract}

\maketitle


\section{ Introduction}

The model of a directed polymer in a random medium 
 plays the role of a `baby spin
glass' model in the field of disordered systems
\cite{Hal_Zha,Der_Spo,Der,Mez,Fis_Hus}.  At low
temperature, there exists a disorder dominated phase, where the 
order parameter is an `overlap' \cite{Der_Spo,Mez,Car_Hu,Com}.
In finite dimensions, a scaling droplet theory was proposed
 \cite{Fis_Hus,Hwa_Fis},
in direct correspondence with the droplet
 theory of spin-glasses \cite{Fis_Hus_SG},
whereas in the mean-field version of the model on the Cayley,
a freezing transition very similar to the one occurring
in the Random Energy Model was found \cite{Der_Spo}.
The phase diagram as a function of space dimension $d$ is the
following \cite{Hal_Zha}. In dimension $d \leq 2$, there is no free phase,
i.e. any initial disorder drives the polymer into the strong disorder phase,
whereas for $d>2$, 
there exists a phase transition between
the low temperature disorder dominated phase
and a free phase at high temperature  \cite{Imb_Spe,Coo_Der},
where the free energy has its annealed value.
 This phase transition  has been studied exactly on a Cayley tree \cite{Der_Spo}
and on hierarchical lattice \cite{Der_Gri}.
  In finite dimensions, bounds on the critical temperature
$T_c$ have been derived \cite{Coo_Der,Der_Gol,Der_Eva} :
$T_0(d) \le T_c \le T_2(d)$.
The upper bound $T_2(d)$ corresponds to the temperature above which the ratio
$\overline{Z_L^2}/(\overline{Z_L})^2$ remains finite as $L \to
\infty$. The lower bound $T_0$ corresponds to the temperature below which
the annealed entropy becomes negative.
In $d=3$, the critical properties have been studied numerically by
\cite{Der_Gol,Ki_Br_Mo}, with different conclusions.
The study of \cite{Der_Gol} gives a slightly negative value
$\alpha  \simeq -0.1 $, whereas the work of \cite{Ki_Br_Mo}
yields a correlation length
exponent $\nu \sim 4 $, corresponding through hyperscaling, to
$\alpha =2-\nu \sim -2 $.

In related growth models belonging to the KPZ universality class,
 numerical studies  and theoretical arguments \cite{Fo_Ta,Ta_Na_Fo,Do_Ko}
 have found that the height-height correlation function
 is logarithmic at the transition. For the directed polymer model at criticality,
this translates  into logarithmic free energy fluctuations
  $\Delta F(l) \sim (\ln l)^{\sigma}$ with an exponent $\sigma=1/2$
  that has been measured in $d=3$ \cite{Fo_Ta,Ki_Br_Mo}.
  In this paper, we make a droplet analysis at criticality
  based on this logarithmic scaling, in direct correspondence
  with the Fisher-Huse droplet theory of the low temperature phase
  based on the free energy scaling $\Delta F(l) \sim l^{\theta}$.
  The matching between the two droplet distributions below $T_c$ and at $T_c$
  allows us to derive that the typical correlation length $\xi(T)$ of
the low temperature phase 
does not follow a power-law $(T_c-T)^{-\nu}$, but diverges
instead as $ \ln \xi(T) \sim  (  \ln 1/(T_c-T) )^{1/\sigma} $.
    Moreover, we argue that the logarithmic fluctuations of the free
energy at criticality leads to the conclusion that the critical temperature
$T_c$ actually coincides 
with the upper bound $T_2$ derived by Derrida and coworkers, since $T_2$
 corresponds to the temperature below which the ratio
$\overline{Z_L^2}/(\overline{Z_L})^2$ diverges exponentially in $L$. 

The paper is organized as follows.
In Section \ref{wetting}, we explain on the pure wetting and Poland-Scheraga
model of DNA denaturation how the transition can be analyzed
in terms of the distribution of large loops. The same approach will
then be adapted in the remainder of the paper to analyse the freezing transition
of the directed polymer, using the loop distribution between two independent
copies of the polymer in the same disordered sample. 
In section \ref{DPlow}, we describe the Fisher-Huse droplet theory of the low temperature phase
based on the scaling $\Delta F(l) \sim l^{\theta}$.
In section \ref{DPcriti}, we describe the droplet theory based on the logarithmic scaling of the 
free energy at criticality $\Delta F (l) \sim ( \ln l)^{\sigma}$ with $\sigma=1/2$, 
and we obtain the divergence of the correlation length $\xi(T)$ near $T_c$,
as well as the behavior of the overlap.
In section \ref{tct2}, we argue that the transition temperature $T_c$
coincides with the upper bound $T_2$ derived by Derrida and coworkers,
and discuss why it is not the case in other disordered systems. 
Finally, in section \ref{spinglass}, we discuss the case of spin-glasses
for which the Fisher-Huse droplet theory was initially developed.
If one assumes, by analogy with the directed polymer model, that
the free energy of droplet excitations is logarithmic at $T_c$, one obtains
some unusual behavior for the correlation function at criticality.
Our conclusions are summarized in Section \ref{conclusion}.
In the Appendix \ref{appendix}, we explain in more detail the matching
procedure for correlation function that we use in this paper.

\section{ Pure delocalization transitions : analysis in terms of the loop distribution }

\label{wetting}

\subsection{ Wetting and Poland-Scheraga model of DNA denaturation}

The wetting model is defined by the partition function
\begin{equation}
Z_{wetting}(2L) = \displaystyle \sum_{RW} 
\exp \left( \beta \displaystyle \sum_{1 \leq \alpha \leq N}
 \epsilon_{\alpha}\delta_{z_{2 \alpha},0}  \right) 
 \label{zwetting}
  \end{equation}
where the sum is over 
one-dimensional random walks (RW) of $2L$ steps, starting at $z(0)=0$,
with increments $z(\alpha+1)-z(\alpha)=\pm 1$. The random walk
is constrained to remain  in the upper half plane $z \geq 0$, but
gains an adsorption energy $\epsilon_0$ if $z(\alpha)=0$. 

The Poland-Scheraga (PS) model of DNA denaturation \cite{Pol_Scher} is
closely related to the wetting model. It describes 
the configuration of the two complementary strands as a sequence of
bound segments and open loops. Each loop of length $l$ has a polymeric
entropic weight 
\begin{equation}
{\cal M} (l)
\sim \frac{\mu^l }{l^c }
\label{loopweight}
\end{equation}
whereas each contact has a Boltzmann weight $e^{- \beta \epsilon_0}$.
The partial partition function $Z_{PS}(1,\alpha)$ with bound ends at
monomers $(1)$ and $(\alpha)$ satisfies the simple recursion relation
\begin{equation}
Z_{PS}(1,\alpha)=  e^{-\beta \epsilon_{\alpha} }  
  \sum_{\alpha'=1}^{\alpha-1}   {\cal M}(\alpha-\alpha') Z_{PS}(1,\alpha')
\label{recursion}
\end{equation}
The wetting model (\ref{zwetting}) corresponds to a Poland-Scheraga
model with parameter $\mu=2$ and loop exponent $c=3/2$ (this
exponent comes from the first return distribution of a one dimensional
random walk). These models without disorder are simple to solve
explicitly : they undergo a phase transition between
a localized phase at low temperature, characterized by an extensive
number of contacts, and a delocalized phase at high
temperature. The transition is first order for $c>2$, and second order
for $1<c<2$ with correlation length exponent $\nu=1/(c-1)$. Let us now describe
how this transition can be understood from the point of view of the
loop distribution.

\subsection{ Loop distribution in the low temperature phase } 

For the wetting or Poland-Scheraga model with loop exponent $c$,
the loop distribution normalized to unity 
\begin{eqnarray}
 \int dl P(l,T) =1
  \label{norma1}
\end{eqnarray}
can be explicitly computed \cite{gdcanonique} .
Near $T_c$, it is useful to decompose it into two terms
\begin{eqnarray}
 P(l,T) =  P_{finite}(l,T) +  P_{large}(l,T)
\label{finitelarge}
\end{eqnarray}
The first term represents the statistics of finite loops $l=1,2,..$,
whereas the second contribution concerning large loops $l \gg 1$
follows the scaling form
\begin{eqnarray}
 dl P_{large}(l,T)
=  {\cal N}(T)  \frac{dl}{l} \Phi \left( \frac{l}{\xi(T)} \right)
\label{large}
\end{eqnarray}
Here $\xi(T)$ is the correlation length that diverges at the transition $\xi(T_c)=\infty$, 
and the factor ${\cal N}(T)$
represents the `normalization' of large loops.
In Poland-Scheraga model, the dependence in $l$ of the probability
of a large loop of length $l$ involves the entropic weight $1/l^c$
of a free loop of length $l$ entering the definition of the model (\ref{loopweight})
and the extensive free energy cost $\Delta F(l)= 
\frac{l}{\xi(T)}$ (where we have used hyperscaling)
\begin{eqnarray}
  P_{large}(l,T) \sim \frac{1}{l^c} e^{- \beta \frac{l}{\xi(T)} }
\label{factor}
\end{eqnarray}
so that the scaling function $\Phi$ in (\ref{large}) reads
\begin{eqnarray}
  \Phi(\lambda) \sim \frac{1}{\lambda^{c-1}} e^{- \beta \lambda }
\end{eqnarray}

Let us now consider the number of contacts $n_L(T)$.
In the low temperature phase, it is extensive and 
simply reads
\begin{eqnarray}
 n_L(T)= \frac{L}{< b >_T + <l>_T}
\label{contact}
\end{eqnarray}
where $<b>_T$ represents the averaged length of sequences of
consecutive bound monomers (which remains finite as $T \to T_c$), and where
$<l>_T$ represents the averaged 
loop length of the full distribution (\ref{finitelarge})
\begin{eqnarray}
<l>_T = \int dl l \left[ P_{finite}(l,T) +  P_{large}(l,T) \right] = finite
 + \int dl  {\cal N}(T)  \Phi \left( \frac{l}{\xi(T)} \right)
= finite + {\cal N}(T) \xi(T) \int d\lambda \Phi(\lambda)
\end{eqnarray}
For $1<c<2$, this averaged loop length diverges as 
\begin{eqnarray}
<l>_T \sim {\cal N}(T) \xi(T) 
\end{eqnarray}
In wetting and Poland-Scheraga models, the energy is directly proportional
to the number of contacts (\ref{contact}), so that the energy density
also vanishes as
\begin{eqnarray}
e(T)= \frac{1}{<l>_T} \sim \frac{1}{{\cal N}(T) \xi(T) }
\label{exi}
\end{eqnarray}
On the other hand, the singularity of the energy is given by
the derivative of the free energy density $f(T) \sim 1/\xi(T)$ with respect to
temperature. The 
critical behavior of the energy is governed by
\begin{eqnarray}
e(T) \sim - \frac{d}{d T}  \frac{1 }{\xi (T) } \sim
\frac{1}{\xi^2 (T)} \frac{d \xi(T) }{d T}
\label{efree}
\end{eqnarray}
The comparison between Eqs (\ref{exi}) and (\ref{efree}) yields the
following differential equation for the correlation length $\xi(T)$ in
terms of the normalization 
${\cal N}(T)$
\begin{eqnarray}
 \frac{d \ln \xi(T) }{d T} = \frac{1}{{\cal N}(T)  }
\label{equadiff}
\end{eqnarray}

\subsection{ Loop distribution at criticality}

At criticality, the loop distribution is simply given by the entropic weight
that enters the definition of the model (\ref{loopweight})
\begin{eqnarray}
 dl P_{large} (l,T_c) \sim  \frac{dl }{l^c}
\label{largec}
\end{eqnarray}
For $c>2$, the averaged length $<l> = \int dl \ l P_{large} (l,T_c)  $
is finite, so that the number $n(T_c)$ of contacts with the substrate
is still extensive $n(T_c) \sim L$ and the transition is first
order. For $1<c<2$, the averaged length $<l>$ diverges 
\begin{eqnarray}
<l>_{T_c} = \int  dl  l P_{large} (l,T_c) = \infty
\end{eqnarray}
The number of contacts is not extensive anymore at criticality
$n_L(T_c)/L \sim 1/ <l>_{T_c} =0$.
Since the L\'evy sum of $n$ independent variables $l_i$ drawn from the
distribution (\ref{largec}) scales as $l_1+...+l_n \sim n^{1/(c-1)}$,
the number of contacts $n^{pure}_L(T_c)$
in a sample of length $L$ scales as
\begin{equation}
n_L^{pure}(T_c) \sim L^{c-1}
\end{equation}
and the transition is second order. Let us now make the connection
with the low temperature phase. 

\subsection{ Matching the loop distribution in the critical region}

On length of order $l \sim \xi(T)$,
the two expressions (\ref{large}) and (\ref{largec})
for the loop distribution for $T<T_c$ and for $T=T_c$
should be of the same order
\begin{eqnarray}
  P_{large} (\xi(T),T) = P_{large} (\xi(T),T_c)
\label{matching}
\end{eqnarray}
This matching determines
the normalization factor
 ${\cal N}(T)$ in terms of the correlation length $\xi(T)$
\begin{eqnarray}
  {\cal N}(T)  \sim \left( \xi(T) \right)^{1-c}
\label{normawetting}
\end{eqnarray}
So as the transition is approached $T \to T_c$, the correlation length diverges
$\xi(T) \to +\infty$, but the density of these large loops vanishes 
${\cal N}(T) \to 0$.

The differential equation (\ref{equadiff}) becomes
 a closed equation for $\xi(T)$
\begin{eqnarray}
 \frac{d \xi(T) }{d T} \sim \left( \xi(T) \right)^{c}
\end{eqnarray}
The integration with the condition $\xi(T_c)=\infty$ gives
\begin{eqnarray}
T_c-T \sim \int_{\xi(T)}^{+\infty} \frac{dx}{x^c} 
\sim \left( \xi(T) \right)^{1-c}
\end{eqnarray}
leading to 
\begin{eqnarray}
 \xi(T) \sim (T_c-T)^{-\nu} \ \ \ { \rm with} \ \ \nu=\frac{1}{c-1}
\end{eqnarray}
in agreement with the exact solution.
Note that the normalization of large loops vanishes
linearly (\ref{normawetting}) independently of the loop exponent $c$
\begin{eqnarray}
  {\cal N}(T)  \sim \left( \xi(T) \right)^{1-c} \sim (T_c-T)
\end{eqnarray}

\subsection{Important ideas for the following sections on disordered systems}

In this section, we have explained how the delocalization transition
for the pure wetting or Poland-Scheraga models could be interpreted as
a vanishing density ${\cal N}(T \to T_c) \to 0$ of large loops of
characteristic size that diverges $\xi(T \to T_c) \to \infty$. An
important point for the following sections is that the notion of loop
distribution which is natural to consider in the low temperature phase
where the 
number of contacts is extensive $n_L(T) \sim L$, has still a meaning
exactly at $T_c$ even if the number of contacts is not extensive
anymore. It looses its meaning only in the high temperature phase
where the number of contacts remains finite. The main idea of the
following sections in thus that in disordered systems  presenting a
low temperature phase where the order parameter is an overlap, the droplet
distribution is not only useful for $T<T_c$  but has still a meaning
exactly at $T_c$ even if the overlap vanishes. Moreover, it is
precisely this critical droplet distribution that determines the
properties of the transition.

\section{Directed Polymer : loop distribution in the low temperature
phase } 

\label{DPlow}

The directed model is defined by the partition function
over $d$ dimensional random walks $x(i)$ of $L$ steps
\begin{eqnarray}
 Z_L (\beta) = \sum_{\{x(i)\}} e^{- \beta E \left(\{x(i\} \right) } \ \ 
{ \rm with } \ \ E \left(\{x(i\} \right) = \sum_{i=1}^L \epsilon(i,x(i))
\label{defpoldir}
\end{eqnarray}
where the random energies $\epsilon(i,x)$ are independent.
We have already described in the Introduction its main features
with the corresponding references.

\subsection{ Statistics of excitations above the ground state}

The droplet theory for directed polymers \cite{Fis_Hus,Hwa_Fis},
is similar to the droplet theory of spin-glasses \cite{Fis_Hus_SG}.
It is a scaling theory that can be summarized as follows.
At very low temperature $ T \to 0$, all observables are governed by
the statistics of low energy excitations above the ground state.
An excitation of large length $l$ costs a random energy
\begin{eqnarray}
 \Delta E(l) \sim l^{\theta} u
\label{ground}
\end{eqnarray}
where $u$ is a positive random variable distributed with some law $Q_0 (u)$
having  some finite density at the origin  $Q_0 (u=0) >0$.
The exponent $\theta$ is the exponent governing 
the fluctuation of the energy of the ground state
is exactly known in one-dimension
$\theta(d=1)=1/3$ \cite{Hus_Hen_Fis,Kar,Joh,Pra_Spo}
and for the mean-field version on the Cayley tree
 $\theta(d=\infty)=0$ \cite{Der_Spo}.
In finite dimensions $d=2,3,4,5,...$, 
the exponent $\theta(d)$ has been numerically measured
\cite{Tan_For_Wol,Ala_etal,KimetAla,Mar_etal},
and we quote the results of the most precise study
we are aware of \cite{Mar_etal} for dimensions $d=2,3$ :
$\theta(d=2)=0.244$ and $\theta(d=3)  = 0.186$.
Note that the existence of a finite upper critical dimension $d_c$
where the exponent would vanish $\theta(d_c)=0$
has remained a very controversial issue between the numerical studies
\cite{Tan_For_Wol,Ala_etal,KimetAla,Mar_etal}
and various theoretical approaches \cite{Las_Kin,Col_Moo,LeDou_Wie}.

From (\ref{ground}), the probability distribution of 
large excitations $ l \gg 1$ reads within the droplet theory
\begin{eqnarray}
dl \rho (E=0,l)  \sim \frac{ dl }{l} e^{- \beta \Delta E(l)} 
\sim  \frac{ dl }{l} e^{- \beta l^{\theta} u  }
\label{rhodroplet}
\end{eqnarray}
where the factor $dl/l$ comes from the notion of independent excitations
\cite{Fis_Hus}. In particular, its average over the disorder
follows the power-law
\begin{eqnarray}
dl \overline{ \rho (E=0,l) }  
\sim \int_0^{+\infty} du Q_0(u)  \frac{ dl }{l} e^{- \beta l^{\theta} u  }
= T Q(0) \frac{ dl }{l^{1+\theta}}
\label{rhoav}
\end{eqnarray}
This prediction describes very well the numerical data in 
the regime $1 \ll l \ll L$ in dimensions $d=1,2,3$ \cite{DPexcita}.

\subsection{ Low temperature phase governed by a zero-temperature fixed point}

According to the droplet theory, the whole low temperature phase $0<T<T_c$
is governed by a zero-temperature fixed point. 
However, many subtleties arise because the temperature
is actually `dangerously irrelevant'. 
The main conclusions of the droplet analysis \cite{Fis_Hus}
can be summarized as follows.
The scaling (\ref{ground}) governs the free energy cost
of an excitation of length $l$, provided one introduces
a correlation length $\xi(T)$ to rescale the length $l$
\begin{eqnarray}
\Delta F (l ) = \left( \frac{l}{\xi(T) } \right)^{\theta} u
\label{deltaF}
\end{eqnarray}
Here as before, $u$ denotes
 a positive random variable distributed with some law $Q (u)$
having  some finite density at the origin  $Q (u=0) >0$.
Moreover, this droplet free energy is a near cancellation of energy and
entropy contributions that scale as \cite{Fis_Hus}
\begin{eqnarray}
\Delta E (l ) \sim  l^{1/2} w
\label{deltaE}
\end{eqnarray}
where $w$ is a random variable of order $O(1)$ and of zero mean.
The argument is that the energy and entropy are dominated by small
scale contributions 
of random sign \cite{Fis_Hus}, whereas the free energy is optimized
on the coarse-grained scale $\xi(T)$.
These predictions for the energy and entropy have been numerically checked in \cite{Fis_Hus,Wa_Ha_Sc}.

\subsection{Loop distribution for two polymers in the same disordered sample}

We now describe how the analysis of Section \ref{wetting}
for the pure transition of the wetting or Poland-Scheraga models
can be adapted to the present disordered case. 
For $T<T_c$, the number of contacts of two independent polymers
$x(i)$ and $y(i)$
in the same disordered sample 
\begin{eqnarray}
n_L(T) = \sum_{i=1}^L < \delta_{x(i),y(i)}>
\label{ncontact}
\end{eqnarray}
is extensive, and the density of contacts, also called the overlap,
is precisely the order parameter of the low temperature phase
\cite{Der_Spo,Mez,Car_Hu,Com}
\begin{eqnarray}
q(T) = \lim_{L \to \infty} \left( \frac{  n_L(T) }{L} \right)
\label{defoverlap}
\end{eqnarray}
Note that on the Cayley tree where $\theta=0$, the distribution
of this overlap is made of two delta peaks at $q=0$ and at $q=1$ 
\cite{Der_Spo},
whereas in finite dimensions with $\theta>0$, the distribution
of this overlap is expected to be a single delta function
at $\overline{q(T)}$ \cite{Mez}.

One may thus analyse the configuration of two polymers in the same
sample in terms of contacts separated by loops.
Again, it is useful to make the decomposition (\ref{finitelarge})
to concentrate on the contribution of large loops that follows
a scaling form based on the free energy scaling of a droplet
of length $l$ (\ref{deltaF}) 
\begin{eqnarray}
dl P_{large}(l,T) =
 {\cal N}(T)  \frac{dl}{l}
 e^{- \beta \Delta F (l )  }
= {\cal N}(T)  \frac{dl}{l}
 e^{- \beta \left( \frac{l}{\xi(T) } \right)^{\theta} u  }
\label{loop}
\end{eqnarray}
As in the wetting case, the factor $ {\cal N}(T)$ represents the
normalization of large loops that will be determined below. 

The important difference with respect to the pure case is that now
the probability of a large loop at a given spatial position
depends upon a random variable $u$. This introduces
very different behaviors for typical and averaged loop distributions.
The averaged loop distribution has the following
power-law decay in the whole low temperature phase
\begin{eqnarray}
dl \overline{ P_{large}(l) } = \frac{dl}{l} {\cal N}(T) \int_0^{+\infty} du Q (u)
e^{- \beta \left( \frac{l}{\xi(T) } \right)^{\theta} u}
\opsimeq_{ l \to \infty}  {\cal N}(T)  Q (0) T 
\frac{dl}{l}  \left( \frac{\xi(T)}{l } \right)^{\theta} 
\end{eqnarray}
whereas the typical decay is an exponential with exponent $\theta$
\begin{eqnarray}
\overline{ \ln  P_{large}(l) } = \ln ({\cal N}(T) /l) 
 - \beta \left( \frac{l}{\xi(T) } \right)^{\theta} u_0
\end{eqnarray}
with $u_0 = \int_0^{+\infty} du u Q (u)$.
In particular, the typical distribution
\begin{eqnarray}
dl P^{typ} (l) = {\cal N}(T)  \frac{dl}{l}  e^{- \beta  \left( \frac{l}{\xi(T) } \right)^{\theta} u_0 }
\label{ptyp}
\end{eqnarray}
has a finite first moment 
\begin{eqnarray}
<l>_{large} = \int dl l P^{typ} (l)  \sim {\cal N}(T) \xi(T)
\end{eqnarray}
As the transition is approached, this term will govern
the divergence of the first moment of the full droplet distribution 
\begin{eqnarray}
<l>_{tot}= <l>_{finite}+ <l>_{large}  \sim {\cal N}(T) \xi(T)
\end{eqnarray}
As a consequence, the number $n_L(T)$ of contacts  
(and equivalently the number of loops) will have a vanishing density
of order
\begin{eqnarray}
\frac{ n_L(T) }{ L} \sim \frac{1}{ <l>_T} = \frac{1}{ {\cal N}(T) \xi(T) } 
\label{qcontact}
\end{eqnarray}

\subsection{Energy fluctuations near the transition }

Let us now consider the specific heat $c(T)$
that measures the fluctuations of the energy $E_L$
\begin{eqnarray}
T^2 c(T) = \frac{1}{L } < (E_L- <E_L>)^2 >
  = \frac{d^2}{ d \beta^2} \left( \frac{ \ln Z_L(\beta)}{L} \right)
\label{defc} 
\end{eqnarray}
In the high temperature phase, the specific heat coincides with its
annealed value. In particular, for the model (\ref{defpoldir}) 
with the following Gaussian distribution for the 
 random energies $\epsilon(i,x)$
\begin{eqnarray}
\rho(\epsilon) = \frac{1}{\sqrt{2 \pi} } e^{- \frac{\epsilon^2}{2}} 
\end{eqnarray}
the annealed specific heat reads
\begin{eqnarray}
T^2 c_{ann}(T) = 1 
\end{eqnarray}
meaning that the energy fluctuations per monomer are
 $\Delta e^2 =1$.
In the low temperature phase, 
the free energy is above the annealed free energy,
and this implies that the specific heat is less than the annealed value,
with a non-diverging singular part \cite{Der}. More precisely
if one introduces the specific heat exponent $\alpha$, we have
\begin{eqnarray}
c(T)-c_{ann}(T_c) \sim - A (T_c-T)^{-\alpha} \ \ \ {\rm with } \ \ \ A>0 
\ \ {\rm  and } \ \  \alpha \leq 0
\end{eqnarray}

To interpret this loss of energy fluctuations in the low temperature phase
in terms of the droplets, it is convenient to write the 
energy fluctuations in terms of the energy difference between two copies
$x(i)$ and $y(i)$ of the polymer in the same disordered sample
\begin{eqnarray}
&&  \frac{1}{L } < (E_L- <E_L>)^2 >   =
 \frac{1}{2L} < \left(E \left(\{x(i)\} \right)-
E \left(\{y(i)\} \right) \right)^2   > =  \frac{1}{2L} < \left( 
 \sum_{i=1}^L \left[ \epsilon(i,x(i))- \epsilon(i,y(i)) \right]
\right)^2   >\\
&&  =
  \frac{1}{2L} <  
 \sum_{i=1}^L \left[ \epsilon(i,x(i))- \epsilon(i,y(i)) \right]^2
   > +  \frac{1}{L} < 
 \sum_{1 \leq i<j \leq L}\left[ \epsilon(i,x(i))- \epsilon(i,y(i)) \right]
 \left[ \epsilon(j,x(j))- \epsilon(j,y(j)) \right]  >
\end{eqnarray}
In this sum, the monomers $i$ corresponding to a contact $x(i)=y(i)$
do not contribute, whereas the monomers $i$ inside loops $x(i) \neq y(i)$
are expected to have an energy fluctuation of order $\Delta e_i^2 \sim 1$
(Eq \ref{deltaE}). 
As explained above, for $T \geq T_c$ all monomers are characterized
by energy fluctuations $\Delta e^2_i \sim 1$. We thus write  
 that the extensive loss in energy fluctuations just below $T_c$
is proportional to the density of contacts $x(i)=y(i)$
obtained in Eq (\ref{qcontact})
\begin{eqnarray}
 \left[ \frac{1}{L } < (E_L- <E_L>)^2 > \right]_{T_c}
- \left[ \frac{1}{L } < (E_L- <E_L>)^2 > \right]_{T}
\sim  \frac{n_L(T)}{L}
\sim \frac{1}{ {\cal N}(T) \xi(T) } 
\label{fluctE}
\end{eqnarray}

Near $T_c$, the free energy density $ f(T) = -T 
\ln Z_L(\beta) /L $ behaves from hyperscaling as
\begin{eqnarray}
f(T) - f(T_c) =  \frac{1 }{\xi (T) } 
\end{eqnarray}
Deriving twice with respect to temperature (\ref{defc}), 
the singular part of the energy fluctuations behave as
\begin{eqnarray}
 \left[ \frac{1}{L } < (E_L- <E_L>)^2 > \right]_{T_c}
- \left[ \frac{1}{L } < (E_L- <E_L>)^2 > \right]_{T}
 \sim    \frac{d^2}{d T^2}  \left( \frac{1 }{\xi (T) } \right)
\label{chalsing}
\end{eqnarray}

The consistency of Eqs (\ref{fluctE}) and (\ref{chalsing}) then gives
 the following
differential equation for the correlation length $\xi(T)$ in terms of the normalization
${\cal N}(T)$ 
\begin{eqnarray}
\frac{d^2}{d T^2}  \left( \frac{1 }{\xi (T) } \right) \sim  \frac{1}{{\cal N}(T) \xi(T) } 
\label{normaxi}
\end{eqnarray}

\section{ Directed Polymer : loop distribution exactly at criticality }

\label{DPcriti}

\subsection{ Logarithmic fluctuations of the free energy}

Let us now consider what happens for $T=T_c$.
Forrest and Tang \cite{Fo_Ta} have conjectured
from their numerical results on a growth model in the KPZ universality class
and from the exact solution of another model \cite{Blo_Hil}
that the fluctuations of the height of the interface
were logarithmic at criticality.
For the directed polymer model, this translates into
a logarithmic behavior of the free energy fluctuations at $T_c$
\begin{eqnarray}
\Delta F (L,T_c)  \sim (\ln L)^{\sigma} v
\label{fcriti}
\end{eqnarray}
where $v$ is a positive random variable of order one 
distributed with some law $R(v)$, and
where the exponent was measured to be in $d=3$ \cite{Fo_Ta,Ki_Br_Mo}
\begin{eqnarray}
\sigma = \frac{1}{2}
\label{sigma}
\end{eqnarray}

Further theoretical arguments
 in favor of this logarithmic behavior can be 
found in \cite{Ta_Na_Fo,Do_Ko}.
The argument of \cite{Do_Ko} is that the power-law behavior
$  F (L,T_c)  \sim L^{\theta_c}$ is impossible at criticality
so that $\theta_c=0$. From the scaling
relation $\theta_c=2 \zeta_c-1$  between exponents \cite{Hus_Hen},
the roughness exponent $\zeta$ is expected
to be exactly $\zeta_c=1/2$ \cite{Do_Ko}, 
and a renormalization argument then leads to logarithmic
 fluctuations of the free energy \cite{Ta_Na_Fo}.

\subsection{ Loop distribution at criticality  }

With the scaling (\ref{fcriti}), the loop distribution exactly at criticality
reads
\begin{eqnarray}
dl P_{T_c}(l) =  \frac{dl}{l}  e^{- \beta \Delta F(l)  }
=\frac{dl}{l}  e^{- \beta_c (\ln l)^{\sigma} v }
\label{looptc}
\end{eqnarray}
 The averaged loop distribution has then the following
 extremely slow decay 
\begin{eqnarray}
dl \overline{ P_{T_c}(l) } = \frac{dl}{l}  \int_0^{+\infty} dv R (v)
e^{- \beta (\ln l)^{\sigma} v}
\opsimeq_{ l \to \infty}    R (0) T_c 
\frac{dl}{l (\ln l)^{\sigma} }  
\end{eqnarray}
whereas the typical decay is given by
\begin{eqnarray}
\overline{ \ln  P_{T_c}(l) } = - \ln l - \beta_c (\ln l)^{\sigma} v_0
\end{eqnarray}
with $v_0=\int_0^{+\infty} dv v R (v)$.
The typical distribution
\begin{eqnarray}
dl P^{typ} _{T_c}(l) =   \frac{dl}{l}  e^{- \beta (\ln l)^{\sigma}   v_0 }
\label{ptyptc}
\end{eqnarray}
has an infinite  first moment for $0<\sigma<1$ (independently of the prefactor 
$\beta_c v_0$)
\begin{eqnarray}
<l>_{T_c} = \int dl l P^{typ}_{T_c} =\infty
\end{eqnarray}
and the number of contacts $n_L(T_c)$
is not extensive in $L$.

\subsection{ Matching the typical loop distribution in the critical region }

For $l \sim \xi$, the two expressions of the typical correlations
in the low temperature phase (\ref{ptyp}) and at criticality
(\ref{ptyptc}) should coincide as in (\ref{matching})
\begin{eqnarray}
  P_{typ} (\xi(T),T) = P_{typ} (\xi(T),T_c)
\end{eqnarray}
This gives the following relation between
the normalization and the correlation length $\xi(T)$
\begin{eqnarray}
 {\cal N}( T)  \sim  e^{ - K (\ln \xi)^{\sigma}  } 
\label{approx}
\end{eqnarray}
where $K \sim \beta_c v_0$ is some constant.

Using (\ref{normaxi}), we thus obtain the following
closed differential equation for the correlation length $\xi(T)$
\begin{eqnarray}
\frac{d^2}{d T^2}  \left( \frac{1 }{\xi (T) } \right) \sim  \frac{1}{ \xi(T) }    e^{ K (\ln \xi)^{\sigma}    } 
\end{eqnarray}

Near the boundary condition $\xi(T_c)=\infty$,
 the leading divergence of $\xi(T)$ is given by
\begin{eqnarray}
 \frac{1}{ (T_c-T)^2 } \sim    e^{ K (\ln \xi)^{\sigma}    } 
\end{eqnarray}
We thus obtain at leading order that the correlation length $\xi(T)$
diverges as 
\begin{eqnarray}
 \xi (T)   \sim 
 e^{   \left( \frac{2}{K} \ln \frac{1}{T_c-T} \right)^{1/\sigma}+... }
\end{eqnarray}
instead of the usual power-law behavior $\xi(T) \sim (T_c-T)^{-\nu}$.
The free energy difference thus vanishes very slowly as
\begin{eqnarray}
f(T)-f(T_c) \sim \frac{1}{ \xi (T)  } \sim 
 e^{  - \left( \frac{2}{K} \ln \frac{1}{T_c-T} \right)^{1/\sigma}+... }
\end{eqnarray}

The normalization of large droplet vanishes 
at leading order as (\ref{normaxi})
\begin{eqnarray}
{\cal N}(T) \sim (T_c-T)^2
 \end{eqnarray}
and the contact density or overlap as (\ref{defoverlap})
\begin{eqnarray}
q(T) = \lim_{L \to \infty} \left( \frac{  n_L(T) }{L} \right) \sim 
\frac{1}{{\cal N}(T) \xi (T) } \sim 
  e^{ -  \left( \frac{2}{K} \ln \frac{1}{T_c-T} \right)^{1/\sigma}
+ 2 \ln \frac{1}{T_c-T}+... }
\end{eqnarray}

In particular, for the value 
$\sigma=1/2$ measured in $d=3$ \cite{Fo_Ta,Ki_Br_Mo},
we obtain the final results
\begin{eqnarray}
 f(T)-f(T_c) &&  \sim \frac{1}{ \xi (T)  }  \sim 
e^{  -  \left( \frac{2}{K} \ln \frac{1}{T_c-T} \right)^{2} +... }  \\
q(T)  &&   \sim 
  e^{ -  \left( \frac{2}{K} \ln \frac{1}{T_c-T} \right)^{2}
+ 2 \ln \frac{1}{T_c-T}+... }  
\end{eqnarray}

\section{ Directed Polymer : explicit value of the critical temperature  }

\label{tct2}

\subsection{ Exact bounds on $T_c$ derived by Derrida and coworkers} 

Let us first recall the physical meaning of
the exact bounds for the critical temperature derived
by  Derrida and coworkers \cite{Coo_Der,Der_Gol,Der_Eva}
\begin{eqnarray}
T_0(d) \le T_c \le T_2(d)
\label{tcbounds}
\end{eqnarray}
The upper bound $T_2(d)$ corresponds to the temperature above which the ratio
\begin{eqnarray}
R_L(T) = \frac{ \overline{ Z_L^2 } }{ ( \overline{ Z_L })^2 }
\label{ratiodef}
\end{eqnarray}
 remains finite as $L \to \infty$. The lower bound $T_0$ corresponds to the temperature below which
the annealed entropy becomes negative.

In dimensions $d=1,2$, the upper bound is at infinity $T_2=\infty$,
whereas  for $d \geq 3$, the upper bound $T_2$ is finite.
The interpretation is as follows \cite{Der_Eva}. The ratio (\ref{ratiodef})
can be decomposed according to the probability $P_L(m)$ that two
independent usual random walks in dimension $d$ meet $m$ times before
time $L$ 
\begin{eqnarray}
R_L(T) 
= \sum _{m=1}^L P(m) B^m
\end{eqnarray}
where the factor
\begin{eqnarray}
B(T) =\frac{ \overline{ e^{2 \beta \epsilon}  } }{ ( \overline{ e^{ \beta \epsilon} } )^2 }
\end{eqnarray}
can be explicitly computed for any distribution of the site disorder
variable $\epsilon$. In dimensions $d=1,2$, two random walks meet an
infinite number of times as $L \to \infty$, whereas for $d \geq 3$,
they meet a finite number $m$ of times as $L \to \infty$. The
distribution of $m$ decays exponentially 
\begin{eqnarray}
P(m) \sim (1-A) A^m
\end{eqnarray}
where $(1-A)$ is the finite probability of never meeting again.
$T_2$ is defined as the temperature where 
\begin{eqnarray}
A B(T_2)=1
\end{eqnarray}
For $T>T_2$, $B(T)<B(T_2)=1/A$, and the ratio $R_L(T)$ has a finite limit
\begin{eqnarray}
R_{\infty}(T>T_2) = \frac{1-A}{1-AB(T)} 
\end{eqnarray}
For $T<T_2$, $R_L(T)$ is a geometric series of parameter $A B(T) >1$,
and it thus diverges exponentially in $L$
\begin{eqnarray}
R_L(T<T_2) 
\sim (1-A) \sum _{m=1}^L (A B(T) )^m \sim (A B(T) )^L
\end{eqnarray}
Exactly at $T_2$, the ratio diverges but not exponentially
\begin{eqnarray}
R_L(T_2) 
= (1-A) \sum _{m=1}^L 1 \sim L
\label{ratiot2}
\end{eqnarray}

\subsection{ Interpretation  in terms of the probability distribution of free energies }

Let us now interpret the above results of the ratio $R_L(T)$ 
 in terms of the probability distribution $P_L(F)$ of the free energy
$F=- kT \ln Z_L$ over the samples of length $L$. By definition (\ref{ratiodef}),
one has
\begin{eqnarray}
R_L(T) = \frac{  \int dF P_L(F) e^{- 2 \beta F_L}  }{ ( \int dF P_L(F) e^{ - \beta F_L}  )^2 }
\label{ratiofreepdf}
\end{eqnarray}

For $T>T_2(d)$, the ratio $R_{\infty}(T)$ is finite : this means that the fluctuations
of the free energy over the samples
\begin{eqnarray}
\left[ \Delta F_L \right]^2_{samples} =  \int dF P_L(F) F^2 - \left( \int dF P_L(F) F \right)^2
\end{eqnarray}
remain of order  $O(1)$ in the limit $L \to \infty$.

On the other hand, for the directed polymer in the low temperature phase $T<T_c$, 
the fluctuations of free energies over the samples is expected to have the same scaling
as the fluctuations of free energies within the same sample when the
end-point varies 
\cite{Fis_Hus} : the fluctuations of free energy over the samples
are thus governed by the droplet exponent $\theta$
\begin{eqnarray}
\left[ \Delta F_L \right]_{samples} (T<T_c) \sim \left[ \Delta F_L \right]_{droplet} (T<T_c) \sim L^{\theta}
\label{samples-droplets}
\end{eqnarray}
Let us now recall Zhang's argument \cite{Hal_Zha} that allows to determine
the exponent $\eta$ of the tail of the free energy distribution
\begin{eqnarray}
P_L(F \to -\infty) \sim e^{- \left( \frac{ \vert F \vert}{L^{\theta}} \right)^{\eta} } 
\label{taileta}
\end{eqnarray}
Moments of the partition function can be then evaluated by the saddle-point
method, with a saddle value $F^*$ lying in the negative tail (\ref{taileta})
\begin{eqnarray}
\overline{ Z_L^n} = \int dF P_L(F) e^{ - n \beta F_L}  \sim \int dF 
e^{- \left( \frac{ \vert F \vert}{L^{\theta}} \right)^{\eta} } e^{ - n \beta F_L} 
\sim e^{ c(n) L^{ \frac{ \theta \eta}{1-\eta}   } }
\label{saddle}
\end{eqnarray}
Since these moments of the partition function have to diverge exponentially in $L$,
the exponent $\eta$ of the tail (\ref{taileta}) reads in terms of the droplet exponent
\begin{eqnarray}
\eta=\frac{1}{1-\theta} 
\end{eqnarray}

\subsection{ Why $T_c$ coincides with $T_2$ for the directed polymer
 in finite dimensions }

At $T_c$, the fluctuations of the free energy are expected to be logarithmic,
as discussed around Eq (\ref{fcriti})
\begin{eqnarray}
\Delta F \sim (\ln L)^{\sigma}  \ \ \ {\rm with} \ \ \ \sigma=\frac{1}{2}
\end{eqnarray}
From these logarithmic fluctuations, 
 it seems rather difficult to obtain an exponential divergence in $L$ of the ratio $R_L(T_c)$
 (\ref{ratiofreepdf}), so the strict inequality $T_c < T_2$ seems very unlikely.
On the contrary, if $T_c=T_2$, it is very natural to obtain the divergence found 
for the ratio at $T_2$ (\ref{ratiot2})
\begin{eqnarray}
R_L(T_2)  \sim L \sim e^{ \ln L }
\end{eqnarray}
Moreover, to obtain the linear divergence (\ref{ratiot2}), the saddle-point method
described above for the low temperature phase  (\ref{saddle}) gives
that the tail of the free energy distribution should be at criticality
\begin{eqnarray}
P_{T_c}(F \to -\infty) \sim e^{- \left( \frac{ \vert F \vert}{ (\ln L)^{\sigma}} \right)^{\eta_c} } \ \ \ {\rm with } \ \ \eta_c=\frac{1}{1-\sigma} 
\end{eqnarray}
The value $\sigma=1/2$ corresponds to the tail exponent $\eta_c=2$.

Our conclusion is thus that the critical temperature in finite dimension $d$
coincides with the temperature $T_2(d)$ 
\begin{eqnarray}
T_c(d)=T_2(d)
\end{eqnarray}
Explicit expressions for $T_2(d)$ in terms of usual integrals appearing
in the theory of random walks can be found in \cite{Coo_Der,Der_Gol}
for site and bond disorder respectively.

As a final remark, let us mention that for the directed polymer
on the Cayley tree that plays the role of a mean-field version of the model,
the critical temperature does not coincide with the upper bound $T_2$,
but coincides with the lower bound $T_0$ (\ref{tcbounds}) below which
the annealed entropy becomes negative \cite{Coo_Der}
\begin{eqnarray}
T_c(Cayley)=T_0(Cayley) < T_2(Cayley)
\end{eqnarray}
This shows that the mean-field limit of the 
tree structure changes the nature of the transition with
respect to the finite-dimensional case.  
The technical reason seems to be that 
$\theta=0$ and $\Delta F =O(1)$ 
in the low temperature phase of the Cayley tree,
whereas Zhang's argument above is consistent only if $\eta =1/(1-\theta)
>1$ to ensure the convergence in the presence of the exponential term
$e^{- \beta n F}$ (Eq \ref{saddle}). When $\theta=0$, the tail of
$P_L(F \to -\infty)$ is
also an exponential $e^{ a F/(\Delta F)}$ as in the Random Energy Model
\cite{Der_Spo} and one has to take into account the minimal
free energy that can be obtained for a finite size $L$.
From a physical point of view, the reason is that
the configurations of two polymers in the same disordered sample
are very different. In finite dimensions, contacts and loops
alternate extensively, whereas on the tree, the loops simply
do not exist : the two polymers may only coincide over some distance
and then never meet again. Since the exponential tail found for the
Cayley tree actually corresponds to the universal Gumbel tail for the
minimum of independent variables, this shows that the non-exponential
tail found in finite dimensions reflects the importance of
correlations between the free energies of paths due to the presence of
loops.

\subsection{ Why $T_c$ is different from $T_2$ in other finite dimensional
disordered systems  }

The fact that the fluctuations of free energies over the samples have the same scaling
as the droplet excitations within one given sample (\ref{samples-droplets})
is very specific to the directed polymer model.
In other disordered models, such as spin-glasses for instance, the fluctuations
of free energies over the samples scale instead as \cite{We_Ai,Bou_Krz_Mar}.
\begin{eqnarray}
\left[ \Delta F_L \right]_{samples}  \sim  L^{d/2}
\label{clt}
\end{eqnarray}
at any temperature.
This scaling simply reflects the Central-Limit fluctuations of the $L^d$ disorder variables
defining the sample. The directed polymer escapes from these normal fluctuations
because it is a one-dimensional path living in a $1+d$ disordered sample :
each configuration of the polymer only sees $L$ random variables
among the $L^{1+d}$ disorder variables that define the sample,
and the polymer can 'choose' the random variables it sees.
So for other disordered systems having fluctuations over the samples governed by (\ref{clt}),
the ratio $R_L(T)$ will diverge exponentially as any temperature.
The temperature $T_2$ is thus infinite
\begin{eqnarray}
T_2=\infty
\end{eqnarray}
and has nothing to do with any critical temperature.
However, the droplet exponent $\theta$ is expected to govern the
correction to the extensive part of the mean value \cite{Bou_Krz_Mar}
\begin{eqnarray}
\overline{ F_L }  \sim  L^{d} f_0 + L^{\theta} f_1
\label{faveraged}
\end{eqnarray}
It can for instance be measured in the free energy difference
upon a change of boundary conditions that forces the introduction
of some domain wall in the sample \cite{Fis_Hus_SG}.

\section{ spin-glasses} 

\label{spinglass}

In this Section, we briefly mention the analogies and differences
between spin-glasses and the directed polymer model described above.

\subsection{ Fisher-Huse droplet theory of the low temperature phase}

Discussions on the droplets statistics in pure Ising models
can be found in Refs \cite{Abr,MFis,Hu_Fi_pure}.
Here we only summarize very briefly
the Fisher-Huse droplet theory of the spin-glass phase \cite{Fis_Hus_SG}.

The free energy to make an excitation of large size $l$ follows the scaling
\begin{eqnarray}
\Delta F(l,T) \sim  \left( \frac{l}{\xi(T)} \right)^{\theta}  u
\end{eqnarray}
where $u$ is a positive random variable of distribution $Q(u)$.

The truncated correlation function 
\begin{eqnarray}
C(r) \sim  \vert < S_0 S_r> -<S_0> < S_r> \vert
\label{truncated}
\end{eqnarray}
for large distance $r$ is governed by the probability that 
the two points $o$ and $r$ 
belongs to the same large droplet 
\begin{eqnarray}
C(r)  \sim {\cal N}(T) e^{ - \left( \frac{r}{\xi(T)} \right)^{\theta} u }
\label{clow}
\end{eqnarray}
where the prefactor ${\cal N}(T)$
corresponds to the Edwards-Anderson parameter
\begin{eqnarray}
{\cal N}(T) \sim C(\xi(T)) \sim q_{EA}(T) = \frac{1}{V} \sum_i < S_i>^2 
\end{eqnarray}
The typical decay of the correlation function is
\begin{eqnarray}
\overline{ \ln C(r) } = 
\ln q_{EA}(T) - \left( \frac{r}{\xi(T)} \right)^{\theta} u_0
\label{typiquelow}
\end{eqnarray}
whereas the mean-square correlation function
\begin{eqnarray}
\overline{  C^2(r) } \sim T Q(0)  q_{EA}^2(T)
 \left( \frac{\xi(T)}{r} \right)^{\theta} 
\end{eqnarray}
has a power-law decay in the whole low temperature phase.

\subsection{ spin-glasses at criticality}

As mentioned above, the exponent $\theta$
 of excitations is expected to govern
the correction to extensivity of the averaged free energy in the low
temperature  phase
(Eq. \ref{faveraged}). 
At criticality where the extensive part of the free energy vanishes,
the averaged free energy is thus expected to be governed by
the free energy scale of excitations at $T_c$
\begin{eqnarray}
\overline{ F_L }  (T_c) \sim  \left[ \Delta F_L(T_c) \right]_{droplet} 
\label{faveragedtc}
\end{eqnarray}
According to the argument of \cite{Do_Ko}, a power-law behavior
\begin{eqnarray}
\overline{ F_L }  (T_c) \sim  L^{\theta_c}
\end{eqnarray}
is not possible at a critical point, so that $\theta_c=0$.
The droplet free energy at criticality can thus only be logarithmic
\begin{eqnarray}
  \left[ \Delta F_L(T_c) \right]_{droplet}  \sim (\ln L)^{\sigma} v
  \label{freelogtc}
\end{eqnarray}
where $v$ is a random variable. 
In pure models, the power-law decay of truncated correlation
can be interpreted as a logarithmic cost with $\sigma_{pure}=1$.
It therefore seems natural to expect $0<\sigma<1$, since in the low
temperature phase, the free energy cost $l^{\theta}$ in the presence
of disorder is much less than the free energy cost $l^{d-1}$ of pure
systems. 

Then the truncated correlation (\ref{truncated})
is expected to behave at large distance as
\begin{eqnarray}
C(r) \sim e^{- \beta \Delta F_r(T_c)}  \sim 
  e^{ - \beta \left( \ln r  \right)^{\sigma} v }
\end{eqnarray}
The typical decay is then given by
\begin{eqnarray}
\overline{ \ln C(r) } =  - \left( \ln r  \right)^{\sigma} \beta v_0
\label{typiquetc}
\end{eqnarray}
whereas the mean-square correlation function
  decays very slowly as
\begin{eqnarray}
\overline{  C^2(r) } 
\sim \frac{ T Q(0) }{\left( \ln r  \right)^{\sigma}}  
\end{eqnarray}
in contrast with the usual power-law 
correlation $\overline{  C^2(r) } \sim 1/r^{d-2+\eta}$
\cite{Fis_Hus_SG}.
The matching at scale $r \sim \xi(T)$
between the typical correlation functions at $T<T_c$ 
(Eq. \ref{typiquelow}) and at $T_c$ (Eq. Ref \ref{typiquetc} )
 yields the following relation between the Edwards-Anderson parameter
$q_{EA}(T)$ and the correlation length $\xi(T)$
\begin{eqnarray}
\ln q_{EA}(T) \sim  - \left( \ln \xi(T)  \right)^{\sigma}
\end{eqnarray}

For the directed polymer model, it was possible to carry
the discussion further because the one dimensional structure
provides a connection between the divergence of large loops
and the total number of small loops (\ref{qcontact}).
In spin-glasses, the relation between properties of large droplets
that dominate the correlation function and properties of small
droplets that dominate the energy fluctuations depends
on geometric assumptions on the shape of the droplets and on their
spatial organization. This goes beyond the present work.

\section{ Conclusions and perspectives} 

\label{conclusion}

In this paper, we have proposed a coherent
picture of the freezing transition
of directed polymers in dimension $d \geq 3$ 
from the following building blocks \\ 
(i) the Fisher-Huse droplet theory of the low temperature phase \\
(ii) the Forrest-Tang result concerning the height-height
correlation exactly at criticality in related growth models
belonging to the KPZ universality class \\
(iii) the exact bounds on the critical temperature of Derrida
and coworkers.

Our main conclusions are that the critical temperature $T_c$
coincides with the upper bound $T_2$ derived by Derrida
and coworkers, and that the logarithmic fluctuations 
at criticality $\Delta F(l,T_c)
\sim (\ln l)^{\sigma}$ with $\sigma=1/2$
leads to unusual critical properties.
In particular, the typical correlation
length $\xi(T)$ of the low temperature phase, diverges 
as $ \ln \xi(T) \sim ( - \ln (T_c-T) )^{1/\sigma}
\sim ( - \ln (T_c-T) )^{2}  $, 
instead of the usual power-law $\xi(T) \sim
(T_c-T)^{-\nu}$. These results emerge from the following picture.
Below $T_c$, the number of contacts of two polymers in the same
disordered sample is extensive, and their configuration
can be thus described as a sequence of contacts and loops.
As the transition is approached, 
the averaged length loops diverges, but our main point
is that the loop distribution has still a meaning exactly at $T_c$.
(It looses its meaning only in the high temperature phase). Moreover, it is
precisely this critical droplet distribution that determines the
properties of the transition. 

 Finally, since the directed polymer plays the role of a
baby spin-glass model, we have briefly mentioned
some similarities and differences with the directed polymer model.
We have discussed some consequences for
the Edwards-Anderson order parameter and the mean-square
averaged correlation, if one
speculates that the free energy of droplet 
excitations for spin-glasses is also logarithmic at $T_c$. 
 Note that the freezing transition
of the directed polymer model is completely asymmetric w.r.t. $T_c$:
there is no singularity in thermodynamic quantities
as $T \to T_c^+$ \cite{Der}, since the free energy coincides with its
annealed value for $T \geq T_c$. For the spin-glass transition,
this raises the question of the relations between
the critical properties below and above $T_c$.

We are presently studying numerically various aspects
of the freezing transition of the directed polymer in $d=3$
 \cite{future} to see if the scenario proposed in the present paper
can be discriminated from the usual power-law critical behaviors
used previously in the literature to analyse the data
 \cite{Der_Gol,Ki_Br_Mo}.

\appendix

\section{ Matching procedure for correlation functions 
in the critical region}
\label{appendix}

As suggested by the referee, we explain in more details
in this Appendix the matching procedure
used in the text. We illustrate it with explicit examples
concerning the truncated correlation function at large distance
\begin{eqnarray}
C_T(r) =  < S_0 S_r> -<S_0> < S_r>
\label{truncated2}
\end{eqnarray}
in pure spin models.

\subsection{ Finite-size scaling theory for the correlation function}

It is useful to introduce the ratio $\rho = r/\xi(T)$ to
rewrite the correlation function as
\begin{eqnarray}
C_T(r) =  G \left(r,\rho=\frac{r}{\xi(T)}\right)
\end{eqnarray}
At criticality where $\xi(T_c)=\infty$, the correlation function
is given by the limit $\rho \to 0$
\begin{eqnarray}
C_{T_c}(r) =  G \left(r, \rho=0 \right)
\end{eqnarray}
In the opposite regime where $\rho \gg 1$, the correlation
is expected to become a scaling function
 of the ratio $\rho$ with a temperature dependent prefactor ${\cal N}(T)$
\begin{eqnarray}
  C_T(r) \opsimeq_{\rho \gg 1} {\cal N}(T) \Phi \left(\rho =\frac{r}{\xi(T)} \right)
\label{correout}
\end{eqnarray}
The matching procedure between the two regimes $\rho \to 0$ and $\rho \gg 1$
consists in requiring that the two expressions should have the same order of magnitude 
at the matching value $\rho^*=r^*/\xi(T) \sim 1$ : this determines the
normalization factor ${\cal N}(T)$ as 
\begin{eqnarray}
  {\cal N}(T) \sim C_{T_c}(r^* \sim \xi(T))
\label{matchingapp}
\end{eqnarray}
We now illustrate the above scheme with Ising and XY models in $d=2$.

\subsection{The Ising model in $d=2$ }

Exactly at $T_c$, the correlation function is given by the power-law
of exponent $\eta=1/4$
\begin{eqnarray}
C_{T_c}(r)  \simeq \frac{1}{r^{1/4}}
\end{eqnarray}
For $T>T_c$,
the correlation function
 follows the Ornstein-Zernicke form \cite{Hu_Fi_pure}
\begin{eqnarray}
C_{T>T_c}(r)  \simeq {\cal N}^+(T)
 \ \left(\frac{\xi(T)}{r} \right)^{1/2} e^{ - \frac{ r}{\xi(T)} }
\label{cabove}
\end{eqnarray}
 whereas for $T<T_c$
the exact decay  
at large distance is \cite{Wu}
\begin{eqnarray}
C_{T<T_c}(r) \simeq   {\cal N}^-(T)
 \ \left(\frac{\xi(T)}{r} \right)^{2} e^{ - 2 \frac{ r}{\xi(T)} }
\label{cbelow}
\end{eqnarray}
The physical interpretation of this form
in terms of droplets can be found in \cite{Abr,MFis,Hu_Fi_pure}.
The matching relation yields
\begin{eqnarray}
  {\cal N}^{\pm}(T) \sim C_{T_c}(r^* \sim \xi(T)) \sim  \frac{1}{\xi^{1/4}(T)}
\end{eqnarray}
Using $\xi(T) \sim \vert T_c-T \vert^{-1}$, the normalizations of 
(\ref{cabove}) and (\ref{cbelow}) are in agreement with the exact
results of \cite{Wu}.

More generally in spin models in dimension $d$ for $T<T_c$,
 the normalization ${\cal N}^-(T)$
is directly related to the order parameter $m(T)=<S_i(T)>$ via 
\begin{eqnarray}
\label{normam}
  {\cal N}^{-}(T) \sim m^2(T)
\label{norma}
\end{eqnarray}
 The decay of the correlation function at criticality defines the exponent $\eta$
\begin{eqnarray}
C_{T_c}(r) \sim \frac{1}{r^{d-2+\eta}} 
\label{isingtc}
\end{eqnarray}
The matching procedure (\ref{matchingapp}) then reads
\begin{eqnarray}
  {\cal N}(T) \sim C_{T_c}(r^* \sim \xi(T)) \sim  \frac{1}{\xi^{d-2+\eta}(T)}
\label{isingnorma}
\end{eqnarray}
With Eq (\ref{normam}), this matching condition 
simply corresponds to the usual relation
$2 \beta = (d-2+\eta) \nu$ between critical exponents.

\subsection{ The XY model in $d=2$}

Since the correlation length is infinite in the whole low temperature phase
($T \leq T_c$), we consider only the matching procedure for $T >
T_c$. Exactly at $T_c$, the correlation function is given by
\cite{Kos,Ami_Gol_Gri} 
\begin{eqnarray}
C_{T_c}(r)  \simeq \frac{( \ln r)^{1/8}}{r^{1/4}}
\end{eqnarray}
For $T>T_c$,
the correlation function
 follows the Ornstein-Zernicke form \cite{Hei_Pel}
\begin{eqnarray}
C_{T>T_c}(r)  \simeq {\cal N}^+(T)
 \ \left(\frac{\xi(T)}{r} \right)^{1/2} e^{ - \frac{ r}{\xi(T)} }
\label{caboveXY}
\end{eqnarray}
The matching procedure (\ref{matchingapp}) yields 
\begin{eqnarray}
  {\cal N}^+(T) \sim C_{T_c}(r^* \sim \xi(T)) \sim  \frac{( \ln \xi(T))^{1/8}}
{\xi^{1/4}(T)}
\label{matchingappXY}
\end{eqnarray}
where the correlation length presents the essential singularity divergence 
 $\xi(T) \sim e^{A/(T_c-T)^{1/2}}$.

\subsection{ Discussion}

The above examples show that whenever the critical correlation
function is a  pure power-law ($C_{T_c}(r) \sim r^{-\eta}$), the
off-critical correlation function can be rewritten in the factorized form 
\begin{eqnarray}
C_{T}(r)  \simeq C_{T_c}(r) \Psi \left(\rho =\frac{r}{\xi(T)} \right)
\label{factoriz}
\end{eqnarray}
that corresponds to the form (\ref{correout}) in the regime $\rho \gg 1$
 with $\Psi(\rho)= \rho^{d-2+\eta} \Phi(\rho)$ and ${\cal N}(T)=1/\xi^{d-2+\eta}(T)$.

However, if $C_{T_c}(r)$ is not a pure power-law,
as in the XY case , the simple factorization (\ref{factoriz}) cannot
be written.

The relations with the polymer models discussed in the text are as
follows. The loop distribution $P(l,T)$ plays the role of a 
correlation function.

 In the pure Poland-Scheraga model where
the loop distribution at criticality is a power-law (\ref{largec}),
the off-critical loop distribution can be written in the factorized
form (\ref{factor})
\begin{eqnarray}
  P_{large}(l,T) \sim P_{large}(l,T_c) \Psi \left( \frac{l}{\xi(T)}  \right)
\end{eqnarray}
as in the Ising model above. On this form, the limit $\xi(T) \to
\infty$ can be taken to recover the critical distribution.

In the directed polymer model where the critical loop distribution
is not a power-law but involves logarithm (\ref{looptc}), the
off-critical loop distribution (\ref{ptyp}) cannot be written 
in a factorized form involving the critical loop distribution, as in
the XY model discussed above. In particular, on the explicit
expression (\ref{ptyp}) with (\ref{approx}), valid in the regime $l \gg
\xi(T)$, one cannot take blindly the limit $\xi(T) \to \infty$ to
recover the critical loop distribution.


\begin{thebibliography}{99}

\bibitem{Hal_Zha}
T. Halpin-Healy and Y.-C. Zhang, Phys. Repts., {\bf 254}, 215 (1995).



 \bibitem{Der_Spo}
B. Derrida and H. Spohn, J. Stat. Phys., {\bf 51}, 817 (1988).


\bibitem{Der}
B. Derrida, Physica {\bf A163}, 71 (1990).

\bibitem{Mez}
M. M\'ezard, J. Phys. (France), {\bf 51}, 1831 (1990).

\bibitem{Fis_Hus}
D.S. Fisher and D.A. Huse,  Phys. Rev. {\bf B43}, 10728  (1991).



\bibitem{Car_Hu}
P. Carmona  and Y. Hu, Prob. Th. and Rel. Fields 124, 431 (2002);
P. Carmona  and Y. Hu, math.PR/0601670. 

\bibitem{Com}
F. Comets, T. Shiga and N. Yoshida, Bernoulli  9, no. 4, 705 (2003).


\bibitem{Hwa_Fis}
T. Hwa and D.S. Fisher,  Phys. Rev. {\bf B49}, 3136 (1994).



\bibitem{Fis_Hus_SG}
D.S. Fisher and D.A. Huse, Phys. Rev. {\bf B38}, 386 (1988).

\bibitem{Imb_Spe}
J. Z. Imbrie and T. Spencer, J. Stat. Phys. {\bf 52}, 609 (1988).

\bibitem{Coo_Der}
J. Cook and B. Derrida, J. Stat. Phys. {\bf 57}, 89 (1989).

\bibitem{Der_Gri}
B. Derrida and R.B. Griffiths, Europhys. Lett. {\bf 8}, 111 (1989).

\bibitem{Der_Gol}
B. Derrida and O. Golinelli, Phys. Rev. {\bf A41}, 4160 (1990).

\bibitem{Der_Eva}
M. R. Evans and B. Derrida, J. Stat. Phys. 69 , 427 (1992 ).

\bibitem{Ki_Br_Mo}
J.M. Kim, A.J. Bray and M.A. Moore, Phys. Rev. {\bf A44}, R4782 (1991).

\bibitem{Fo_Ta}
B.M. Forrest and L-H. Tang, Phys. Rev. Lett., {\bf 64}, 1405 (1990)

\bibitem{Do_Ko}
C.A. Doty and J.M. Kosterlitz, Phys. Rev. Lett., {\bf 69}, 1979 (1992).

\bibitem{Ta_Na_Fo}
L-H. Tang, T. Nattermann and B.M. Forrest, Phys. Rev. Lett., {\bf 65},
2422 (1990).


\bibitem{Pol_Scher} D. Poland and H.A. Scheraga eds., \textit{Theory of
Helix-Coil transition in Biopolymers}, Academic Press, New York (1970).

\bibitem{gdcanonique}C. Monthus, T. Garel and H. Orland,
Eur. Phys. J., {\bf B17}, 121 (2000).

\bibitem{Hus_Hen_Fis}
D. A. Huse, C. L. Henley, and D. S. Fisher, 
Phys. Rev. Lett. {\bf 55}, 2924 (1985).

\bibitem{Kar}
M. Kardar, Nucl. Phys. {\bf B290}, 582 (1987).

\bibitem{Joh}
K. Johansson, Comm. Math. Phys. {\bf 209}, 437 (2000).
 
\bibitem{Pra_Spo}
 M. Pr\"ahofer and H. Spohn,  Physica {\bf A279}, 342 (2000) ; 
M. Pr\"ahofer and H. Spohn, Phys. Rev. Lett. {\bf 84}, 4882   (2000) ; 
    M. Pr\"ahofer and H. Spohn, J. Stat. Phys. {\bf 108}, 1071 (2002)  ; 
  M. Pr\"ahofer and H. Spohn, J. Stat. Phys. {\bf 115}, 255 (2002).




\bibitem{Tan_For_Wol}
L.H. Tang, B.M. Forrest and D.E. Wolf, Phys. Rev. {\bf A45}, 7162 (1992).

\bibitem{Ala_etal}
T. Ala-Nissila, T. Hjelt, J.M. Kosterlitz and V. Venalainen,
J. Stat. Phys. {\bf 72}, 207 (1993).

\bibitem{KimetAla}
T. Ala-Nissila,  Phys. Rev. Lett. {\bf 80}, 887 (1998);
J.M. Kim, Phys. Rev. Lett. {\bf 80}, 888 (1998).

\bibitem{Mar_etal}
E. Marinari, A. Pagnani and G. Parisi, J Phys. {\bf A33}, 8181 (2000);
E. Marinari, A. Pagnani and G. Parisi and Z. Racz, 
Phys. Rev. {\bf E65}, 026136 (2002).

\bibitem{Las_Kin}
M. Lassig and H. Kinzelbach, Phys. Rev. Lett. {\bf 78}, 903 (1997).

\bibitem{Col_Moo}
F. Colaiori and M. A. Moore, Phys. Rev. Lett. {\bf 86}, 3946 (2001).

\bibitem{LeDou_Wie}
P. Le Doussal and K.J. Wiese, Phys. Rev. {\bf E72}, 035101 (2005).

\bibitem{DPexcita}
C. Monthus and T. Garel, cond-mat/0602200.

\bibitem{Wa_Ha_Sc}
X-H. Wang, S. Havlin and M. Schwartz, J. Phys. Chem. {\bf B104}, 3875
(2000);
X-H. Wang, S. Havlin and M. Schwartz,  Phys. Rev. {\bf E63}, 032601 (2001).


\bibitem{Blo_Hil}
H. W. J Bl\"ote and H.J. Hilhorst, J. Phys. A: Math. Gen. 15, L631 (1982).

\bibitem{Hus_Hen}
D. A. Huse and C. L. Henley,
Phys. Rev. Lett. {\bf 54}, 2708 (1985).

\bibitem{We_Ai}
J. Wehr and M. Aizenman, J. Stat. Phys. 60 (1990) 287.

\bibitem{Bou_Krz_Mar} J.-P. Bouchaud, F. Krzakala and O.C. Martin,
Phys. Rev. {\bf B68}, 224404 (2003). 

\bibitem{Abr}
D.B. Abraham, Phys. Rev. Lett. 50, 291 (1983).

\bibitem{MFis}
M. Fisher, J. Stat. Phys. 34, 667 (1984). 

\bibitem{Hu_Fi_pure}
D.A. Huse and D.S. Fisher, Phys. Rev. B 35, 6841 (1987).

\bibitem{future}
C. Monthus and T. Garel, in preparation.

\bibitem{Wu}
T.T. Wu, Phys. Rev. 149, 380 (1966).

\bibitem{Kos}
J.M. Kosterlitz, J. Phys. C 7 , 1046 (1974).

\bibitem{Ami_Gol_Gri}
D.J. Amit, Y.Y. Goldschmidt and G. Grinstein, J. Phys. A 13 , 585 (1980).

\bibitem{Hei_Pel}
S.W. Heinekamp and R.A. Pelcovits, Phys. Rev. B 32, 4528 (1985).


\end{thebibliography}
\end{document}